\documentclass[a4paper,fleqn,usenatbib]{mnras}
\usepackage{graphicx}
\usepackage{epsfig}
\usepackage{amsmath}

\usepackage[T1]{fontenc}
\usepackage{ae,aecompl}
\usepackage{multicol}
\usepackage{booktabs}
\usepackage{amssymb}	
\usepackage{float}

\newcommand       \mum        {\,{\rm \mu m}}
\newcommand       \simali       {{\sim}\,}

\newcommand	\gtsim	{\lower.5ex\hbox{$\buildrel > \over \sim$}}
\newcommand     \ltsim  {\lower.5ex\hbox{$\buildrel < \over \sim$}}
\newcommand	\simgt	{\lower.5ex\hbox{$\buildrel > \over \sim$}}
\newcommand     \simlt  {\lower.5ex\hbox{$\buildrel < \over \sim$}}

\newcommand       \yr           {\,{\rm yr}}

\title[Chemical differentiation across grain sizes]
{Interstellar chemical differentiation across grain sizes}

\author[Ge, He \& Li]
{ J. X. Ge$^{1,2}$\thanks{E-mail: gejixing@ynao.ac.cn },
  J. H. He$^{1,3}$\thanks{E-mail: jinhuahe@ynao.ac.cn },
  and Aigen Li$^{4}$\\
$^{1}$ Key Laboratory for the Structure and Evolution of Celestial Objects, Yunnan Observatories, Chinese Academy of Sciences, \\ P.O. Box 110, Kunming, 650011, Yunnan Province, China\\
$^{2}$ University of the Chinese Academy of Sciences, Yuquan Road 19, Shijingshan Block, 100049, Beijing, China\\
$^{3}$ Chinese Academy of Sciences, South America Center for Astrophysics (CASSACA) at Cerro Cal$\acute{\mathrm a}$n, \\
Camino El Observatorio 1515, Las Condes, Santiago, Chile \\
$^{4}$ Department of Physics and Astronomy, University of Missouri, Columbia, MO 65211, USA
}

\date{Accepted XXX. Received YYY; in original form ZZZ}

\pubyear{2016}

\begin{document}
\label{firstpage}
\pagerange{\pageref{firstpage}--\pageref{lastpage}}
\maketitle

\begin{abstract}
In this work we investigate the effects of ion accretion and size-dependent dust temperatures on the abundances of both gas-phase and grain-surface species. While past work has assumed a constant areal density for icy species, we show that this assumption is invalid and the chemical differentiation over grain sizes are significant. We use a gas-grain chemical code to numerically demonstrate this in two typical interstellar conditions: dark cloud (DC) and cold neutral medium (CNM). It is shown that, although the grain size distribution variation (but with the total grain surface area unchanged) has little effect on the gas-phase abundances, it can alter the abundances of some surface species by factors up to $\simali$2--4 orders of magnitude. The areal densities of ice species are larger on smaller grains in the DC model as the consequence of ion accretion. However, the surface areal density evolution tracks are more complex in the CNM model due to the combined effects of ion accretion and dust temperature variation.
The surface areal density differences between the smallest ($\sim0.01\mu$m) and the biggest ($\sim0.2\mu$m) grains can reach $\simali$1 and $\simali$5 orders of magnitude in the DC and CNM models, respectively. 

\end{abstract}

\begin{keywords}
astrochemistry -- ISM: abundances -- ISM: clouds -- (ISM:) dust
\end{keywords}



\section{Introduction}

Generally, gas-grain chemical models assume a uniform, single grain size ($a=0.1\mum$) to represent the grain surface \citep[e.g.][]{hase1992,garr2006,garr2008,seme2010,garr2013a,garr2013b}. In this context, many advancements have been made to better interpret observations, e.g., the modified-rate method \citep[e.g.][]{Case1998,Stan2001,garrod2008}, and Monte Carlo simulations of coupled gas-grain chemistry \citep[e.g.][]{vasy2009,vasy2013,garr2013b,chang2014}. However, some studies have considered the effects of grain size distributions and dust temperature variation on chemistry \citep{Cuppen2006,acha2011,Pauly2016}. These studies considered accretion processes for neutral species but ignored ion accretion, which may be an important contributor to surface material. A proper treatment of grain size distribution is becoming a routine practice in astrochemical modeling. Following the treatment of ion accretion described in our previous work \citep[][hereafter GHY]{Ge2016}, in this Letter the coupling of the grain size distributions with ion accretion processes and dust temperature distributions is analysed and simulated to illustrate its important chemical consequences.

\section{Models and Method}
\label{method}

\subsection{Physical models}
\label{phy_model}

 In order to test the effects of ion accretion, we adopt two typical physical models that represent two extreme conditions in the cold interstellar medium: dark cloud (DC) and cold neutral medium (CNM).
For the DC model, the gas kinetic temperature is $T_{\rm gas}=10.0$\,K, gas density is $n_{\rm H}=10^4$ cm$^{-3}$ and the visual extinction is taken to be $A_{\rm V}=7.5$\,mag. For the CNM model, $T_{\rm gas}=100$\,K, $n_{\rm H}=30$ cm$^{-3}$ and $A_{\rm V}=0$\,mag. The parameters of dust will be discussed in detail below. The unattenuated far ultraviolet (FUV) flux $\chi$ is fixed to the standard interstellar radiation field (ISRF) strength \citep{drai1978} for both DC and CNM models.

\subsection{Dust model}
\label{gsd}
\begin{table}
\centering
\begin{minipage}{80mm}
\caption{Dust model parameters (see Section~\ref{gsd}).}
\label{tab_dust}
\begin{tabular}{@{}cc@{  }ccc@{  }c@{}}
\hline
\multicolumn{3}{|c|}{DC \& HY} & \multicolumn{3}{|c|}{CNM \& WD} \\ \cmidrule(r){1-3}\cmidrule(r){4-6}
$a$  & $n_{\rm d}(a)$ & $T_{\rm d}(a)$  & a  & $n_{\rm d}(a)$ & $T_{\rm d}(a)$   \\
 (cm) &(cm$^{-3}$)  & K  & (cm)  & (cm$^{-3}$) & K   \\
\hline
1.13e-06 & 1.17e-07 & 9.1 & 1.15e-06 & 3.33e-09 & 17.9 \\
1.37e-06 & 7.87e-08 & 9.1 & 1.55e-06 & 1.84e-09 & 17.9 \\
1.57e-06 & 5.96e-08 & 9.1 & 2.13e-06 & 9.78e-10 & 17.8 \\
1.81e-06 & 4.49e-08 & 9.1 & 2.97e-06 & 5.01e-10 & 17.5 \\
2.28e-06 & 2.85e-08 & 9.1 & 4.23e-06 & 2.48e-10 & 16.9 \\
3.18e-06 & 1.47e-08 & 9.1 & 6.11e-06 & 1.19e-10 & 16.2 \\
4.75e-06 & 6.54e-09 & 9.1 & 8.96e-06 & 5.53e-11 & 15.6 \\
7.90e-06 & 2.37e-09 & 9.1 & 1.32e-05 & 2.54e-11 & 15.2 \\
1.57e-05 & 5.97e-10 & 9.1 & 1.96e-05 & 1.16e-11 & 14.9 \\
\hline
\end{tabular}\\ [1mm]
\end{minipage}
\end{table}
To investigate the chemical differentiation over grain sizes, we adopt different grain size distributions for the DC and CNM models. The shattering and coagulation calculation of dust grains in a turbulent interstellar medium of \citet{hira2009} (hereafter HY) shows that small grains can grow to bigger ones by coagulation in DC due to the very small grain motion and low temperature. Therefore, we adopt the simulated silicate grain size distribution at 5\,Myr model age of HY for our DC model. For the CNM model, we adopt the grain size distribution from \citet{wein2001} (hereafter WD) with $R_{\rm V}=3.1$ and $b_{\rm c}=6\times 10^{-5}$.

Unlike the assumptions used in previous gas-grain models \citep[e.g.][]{hase1992,seme2010,acha2011}, the dust temperature should be different from the gas temperature and vary over grain sizes \citep[see e.g.][]{Li2012}. For the CNM model, we adopt the dust temperature profile over grain sizes from \citet{Li2012} (the curve with ISRF strength $\chi_{\rm MMP}=1.0$ in their Fig.\,3). For the grain temperatures in our DC model, a trial thermal balance computation with the attenuated interstellar UV radiation as the sole heating source shows decreasing temperature toward small grain size. However, the small-grain temperature may not be so low if cosmic ray heating and UV photon scattering in turbulent clouds are taken into account. Therefore, we adopt a constant grain temperature of 9.1\,K \citep[from a new calculation according to][]{Li2012} for all grain sizes in the DC model. As a result, we only investigate the chemical effects of varying grain temperatures in the CNM model, but not in the DC model.

To check the effects of different grain size distributions, we also adopt a single grain size model ($a=10^{-5}$\,cm) with fixed typical dust-to-gas mass ratio of 0.01 and dust temperatures of 9.1\,K and 18\,K for the DC and CNM models respectively. The gas parameters are set to be the same as in Section~\ref{phy_model}. To exclude the chemical effects of varying total grain surface areas, we set the total grain surface areas of our new models to be the same as that of the single grain-size models. This determines the total amount of dust; thus, the dust-to-gas mass ratio can be different among the models with different grain size distributions. We sample the original grain size distributions with nine grain sizes. Integrated grain surface area is computed as a function of grain size ($0.01\mu {\rm m}\le a\le 0.25\mu {\rm m}$) and homogeneously divided into nine subranges. Then the grain size corresponding to the median grain surface area in each subrange is adopted and the number density of the grains is determined from the conservation of grain surface area within each subrange. The resulting dust grain radii, number density and temperature are listed in Table~\ref{tab_dust}.

\subsection{Chemical model}
\label{reduced_network}
\begin{figure}
\centering
\includegraphics[scale=0.6]{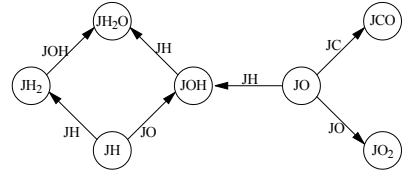}
\caption{The reduced surface reaction network.}
\label{network}
\end{figure}
We adopt the full gas-phase network from \citet{seme2010}. For the surface chemical network, because we will differentiate the chemistry on grains of different sizes, we reduce the surface network to a simpler one that includes only the surface reactions
of ice species composed of H, C and O (see the reduced surface
reaction network in Fig.~\ref{network}) to avoid huge computation.
The (neutral and ion) accretion and desorption for all surface
species in our reduced network are included. We also consider the photo-electron (PE) ejection processes
of grains and a wide enough range of the grain charges from  $-5{\rm e}$ to $+30{\rm e}$
to simulate the variation of grain charges. However, we do not differentiate the chemical abundances on grains with
different charges to reduce computational complexity. Actually, the fast variation of grain charges caused by the PE
and charge accretion (especially the fast electron recombination) may result in efficient chemical mixing
among grains with different charges. This network is also applied to the single grain-size models. 

Our FORTRAN code `ggchem' (see GHY) is modified to model the grain size distribution and the variation of dust temperature. A modified rate equation method similar to \citet{garrod2008} is applied to handle the surface chemistry on small grains. The initial gas abundances are listed in Table~\ref{tab_iabun} and the grains are all bare at the beginning.
\begin{table}
 \centering
 \begin{minipage}{80mm}
  \caption{Initial fractional abundances.}
  \label{tab_iabun}
  \begin{tabular}{llllll}
  \hline
   species & $n(i)/n_{\rm H}$ $^*$ & species & $n(i)/n_{\rm H}$ $^*$ \\
   \hline
   H$_2^{**}$  & 5.0(-1) & Si$^+$ & 8.0(-9) \\
   He     & 9.0(-2)      & Fe$^+$ & 3.0(-9) \\
   C$^+$  & 1.2(-4)      & Na$^+$ & 2.0(-9) \\
   N      & 7.6(-5)      & Mg$^+$ & 7.0(-9) \\
   O      & 2.56(-4)     & P$^+$  & 2.0(-10)\\
   S$^+$  & 8.0(-8)      & CL$^+$ & 1.0(-9) \\
\hline
\end{tabular} \\ [1mm]
Note: Values are taken from \citet{seme2010}.\\
{\em *} a(b)=$a\times 10^b$.\\
{\em **} For the CNM model, all the hydrogen nuclei
         are initially atomic (i.e., $n({\rm H})/n_{\rm H}=1$). 
\end{minipage}
\end{table}
\subsection{Chemical differentiation over grain sizes}
\label{eq_kine}
With the inclusion of ion accretion and grain temperature variation, the chemical kinetic equation describing the formation and destruction of a surface species $i$ on grains with radius $a$ is given as
\begin{equation}
\label{eq3}
\begin{split}
\frac{dn_i^s(a)}{dt} =
 &   \sum_{l,m} \frac{P(r_l\left[T_{\rm d}\right]+r_m\left[T_{\rm d}\right])}{4\pi a^2 n_{\rm d}(a) N_s}n_l^s(a)n_m^s(a) \\
 & - \sum_{i,l} \frac{P(r_i\left[T_{\rm d}\right]+r_l\left[T_{\rm d}\right])}{4\pi a^2 n_{\rm d}(a) N_s}n_i^s(a)n_l^s(a)  \\
 & - k_i^{\mathrm{des}}\left[T_{\rm d}(a)\right]n_i^s(a)\\ 
 & + S_{\rm neu}\pi a^2 V_t(i) n_i^{\mathrm{neu}}n_{\rm d}(a) \\
 & + S_{\rm ion} \pi a^2  \sum_{j} V_t(j) n_j^{\mathrm{ion}} \sum_{Z} \tilde{J} n_{\rm d}(a,Z) ~~,
\end{split}
\end{equation}
where the first and second terms on the right side of the equation are the production and consumption rates of chemical reactions respectively, where $P$ is the reaction probability, $n_{\rm d}(a)$ is the number density of grains with radius $a$, $r_{x}\left[T_{\rm d}\right]=\nu(x)\exp\left[-E_{\rm diff}(x)/T_{\rm d}\right]$ is the thermal diffusion rate of surface species $x$ ($x=i,l,\,{\rm or}\,m$), with $\nu(x)=\sqrt{\frac{2N_sk_bE_{\rm des}(x)}{\pi^2m_x}}$ being the characteristic vibrational frequency in which $N_s$ is the adsorption site density on grains, $E_{\rm des}(x)$ and $E_{\rm diff}(x)$ are the desorption and diffusion energy respectively, $k_b$ is the Boltzman constant; the third term is for the desorption processes including thermal and cosmic-ray induced desorption, which is described by
$k_i^{\rm des}\left[T_{\rm d}(a)\right]=\nu(i)\exp\left[-\frac{E_{\rm des}(i)}{T_d(a)}\right] + f\nu(i)\exp\left[-\frac{E_{\rm des}(i)}{70\,{\rm K}}\right]$ where $f=3.0\times 10^{-19}$ is the ratio of grain cooling timescale via desorption of molecules to the timescale of subsequent heating events; the fourth term is for the neutral accretion process, where $V_{\rm t}(i)=\sqrt{\frac{8k_{\rm b}T_{\rm gas}}{\pi m_i}}$ is the average thermal speed of gas-phase species $i$ with mass $m_i$, $S_{\rm neu}=1.0$ is the sticking coefficient for neutral species, $n_i^{\rm neu}$ is the number density of species $i$; the last term is for the accretion of ionic species $j$ with the Coulomb factor $\tilde{J}$ which is a function of grain size, gas temperature and charge ratio between the grains and the accreted ions (see details in GHY). $S_{\rm ion}=0.5$ is the sticking coefficient of neutral products of ion accretion ( equivalent to assuming that half of the neutral products remain on the grain surface). Note that we sum over the grain charges in the last term because we do not differentiate the surface abundances over grain charges. For more details see \citet{drai1987}, \citet{seme2010} and GHY.

By dividing $4\pi a^2 n_d(a)$ to each term of equation (\ref{eq3}), we finally get the kinetic equation about areal density $\alpha_i^s=n_i^s(a)/\left[ 4\pi a^2 n_{\rm d}(a) \right]$ as
\begin{equation}
\label{eq4}
\begin{split}
\frac{d\alpha_i^s(a)}{dt} = & \sum_{l,m} P(r_l\left[T_{\rm d}\right]+r_m\left[T_{\rm d}\right])\alpha_l^s(a)\alpha_m^s(a)/N_s \\
 & - \sum_{i,l}P(r_i\left[T_{\rm d}\right]+r_l\left[T_{\rm d}\right])\alpha_i^s(a)\alpha_l^s(a)/N_s  \\
 & - k_i^{\mathrm{des}}\left[T_{\rm d}(a)\right]\alpha_i^s(a)\\
 & + S_{\rm neu} V_t(i) n_i^{\mathrm{neu}}/4 \\
 & + \sum_{j} S_{\rm ion} V_t(j) n_j^{\mathrm{ion}} \langle \tilde{J}\rangle/4.
\end{split}
\end{equation}
where $\langle \tilde{J}\rangle = \sum_{Z} \tilde{J} n_{\rm d}(a,Z)/n_{\rm d}(a)$ is defined as the grain charge-weighted average Coulomb factor. From this equation, one can see that all but the neutral accretion process (the fourth term) are dependent on the grain size and/or temperature, which will result in the chemical differentiation over grain sizes. When the product of ion abundance and $\langle \tilde{J}\rangle$ is large enough, we expect that the ion accretion will become important so as to produce significant chemical differentiation over grain sizes. As the dust temperature decreases with increasing dust grain sizes in our CNM model, the thermal desorption rate coefficients and surface reaction rates increase exponentially with increasing grain temperature to reduce the abundances of light surface species (with lower desorption and diffusion energy barriers) on smaller grains much more than on big grains in this model.
In these cases, the usual assumption of constant surface density over grain sizes \citep[e.g.][]{acha2011} becomes invalid.

\section{Simulation results}
\label{Re_and_Di}

\subsection{Comparison with single grain size chemical models}
\label{comp_with_std}
\begin{figure}
\centering
\includegraphics[scale=0.5]{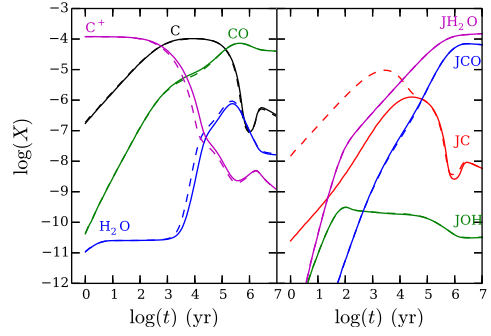}
\caption{The abundance evolution tracks of selected gas-phase (left panel) and surface (right panel) species in the DC model. The solid and dashed lines are used to differentiate the results from single grain-size model and multiple grain-size model}, respectively.
\label{dc_abun}
\end{figure}
\begin{figure}
\centering
\includegraphics[scale=0.5]{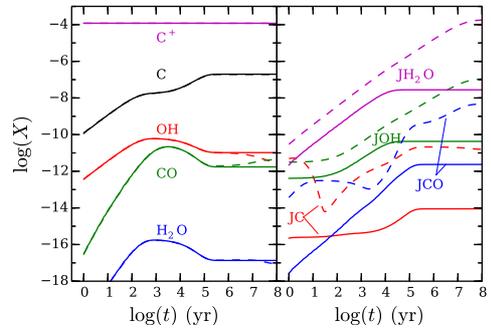}
\caption{Same as Fig.~\ref{dc_abun}, but for the CNM model.}
\label{cnm_abun}
\end{figure}
To check the chemical effects of different grain size distributions, we plot the abundance evolution tracks of some selected gas-phase and surface species calculated from models with a single grain size and with a grain size distribution in Fig.~\ref{dc_abun} and \ref{cnm_abun} for the DC and CNM models, respectively.
From these two figures, we see that the abundance differences of gas-phase species are small (with factors less than $\simali$10) because the gas-phase ions are insensitive to their small loss via ion accretion. However, significant differences occur to some surface species (see right panels in the two figures). For example, in Fig.~\ref{dc_abun}, the surface species JC (red dashed lines) in our DC model with a grain size distribution show an enhanced abundance by factors more than 2 orders of magnitude at $t<10^4$ yr, compared to that in the single grain-size model. For surface species in our CNM model with a grain size distribution (Fig.~\ref{cnm_abun}), the JC, JCO, JOH and JH$_2$O similarly present enhanced abundances by factors up to 4 orders of magnitude in the whole evolution. 

\subsection{The areal density evolution of surface species}
\label{effects}
\begin{figure}
\centering
\includegraphics[scale=0.83]{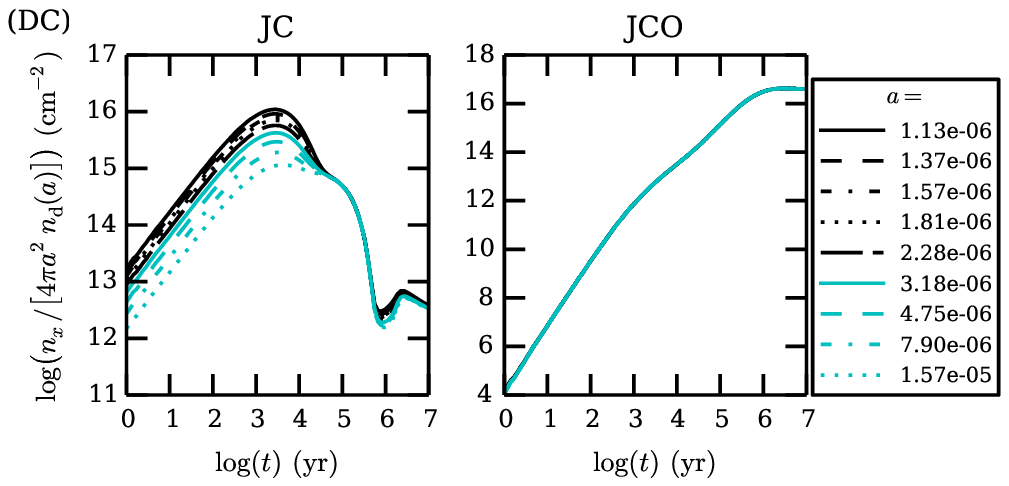}
\includegraphics[scale=0.83]{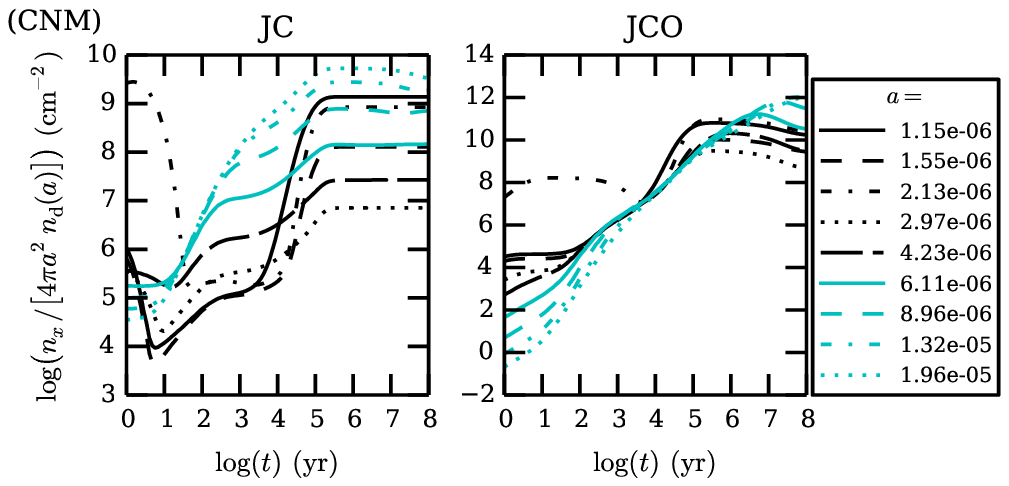}
\caption{Areal density evolution tracks of selected surface species JC (left panels) and JCO (right panels) on grains with different sizes. The panels in the upper and lower rows are for the DC and CNM models, respectively. The combination of line types and colors are used to differentiate the grain sizes.}
\label{dc_cnm_i_density}
\end{figure}
We take JC and JCO as examples since JC shows the largest differentiation of areal density over grain sizes and JCO is a common ice species tightly related to JC.
Their areal density evolution tracks are plotted in Fig.~\ref{dc_cnm_i_density}. The panels in upper and lower rows are for the DC and CNM models, respectively.
For JC in the DC model (upper left panel), the areal densities are larger on smaller grains than on bigger ones by factors more than 1 order of magnitude at $t<10^4\yr$. However, the areal densities of JCO (upper right panel) are nearly the same on all grains because the neutral accretion and desorption are always the leading processes. In the CNM model, the situation is more complex and the areal densities are no longer the smallest on the biggest grains. The areal density differences are also much larger than that in the DC model, reaching  $\sim 5$ orders of magnitude. The reason for the complexity is the interplay between the ion accretion and the variation of dust temperature. Our trial computation (not shown) without dust temperature variation shows that the areal densities also monotonically decrease toward larger grains in the CNM model.

\section{Discussion}
\begin{figure}
\centering
\includegraphics[scale=0.43]{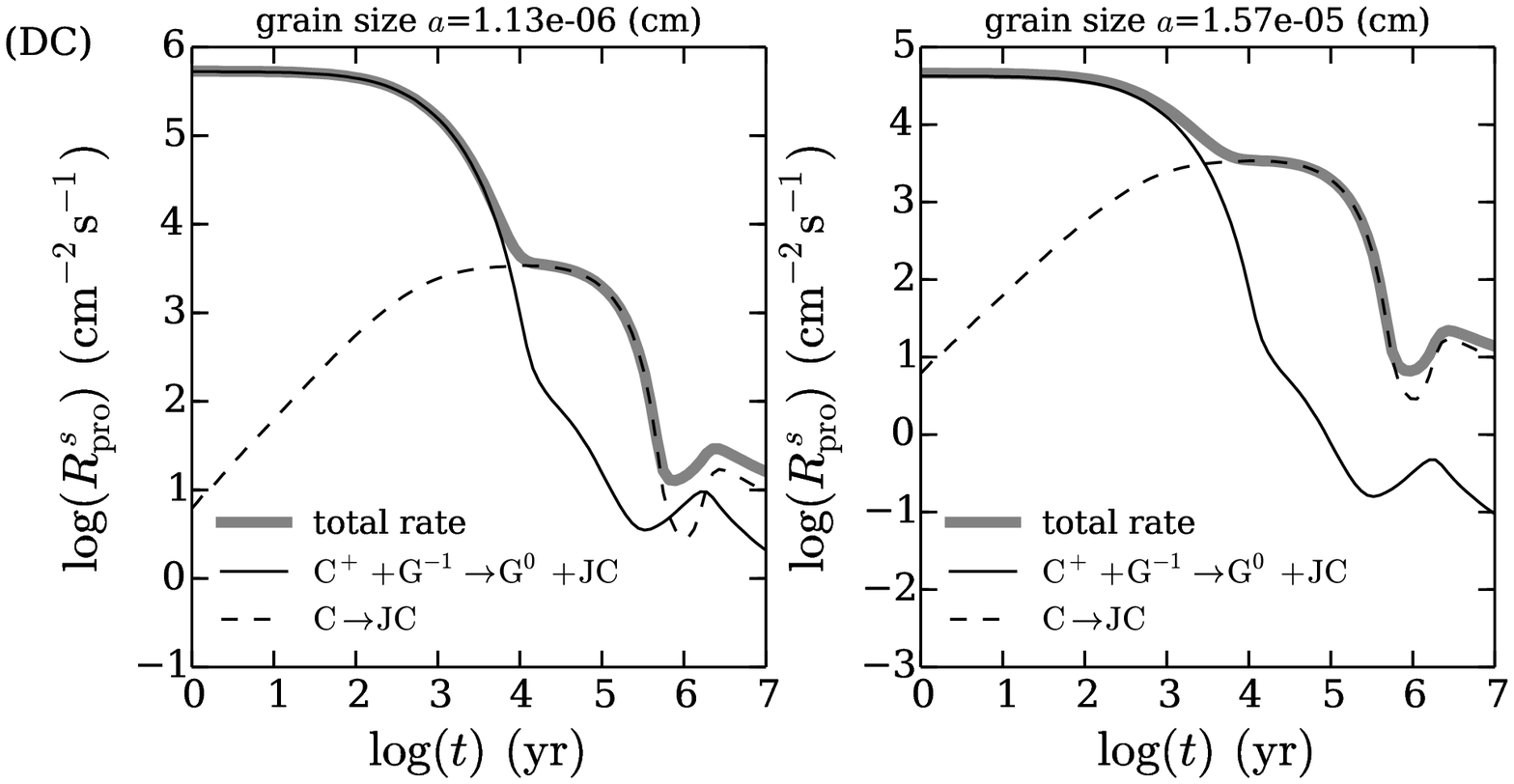}
\includegraphics[scale=0.43]{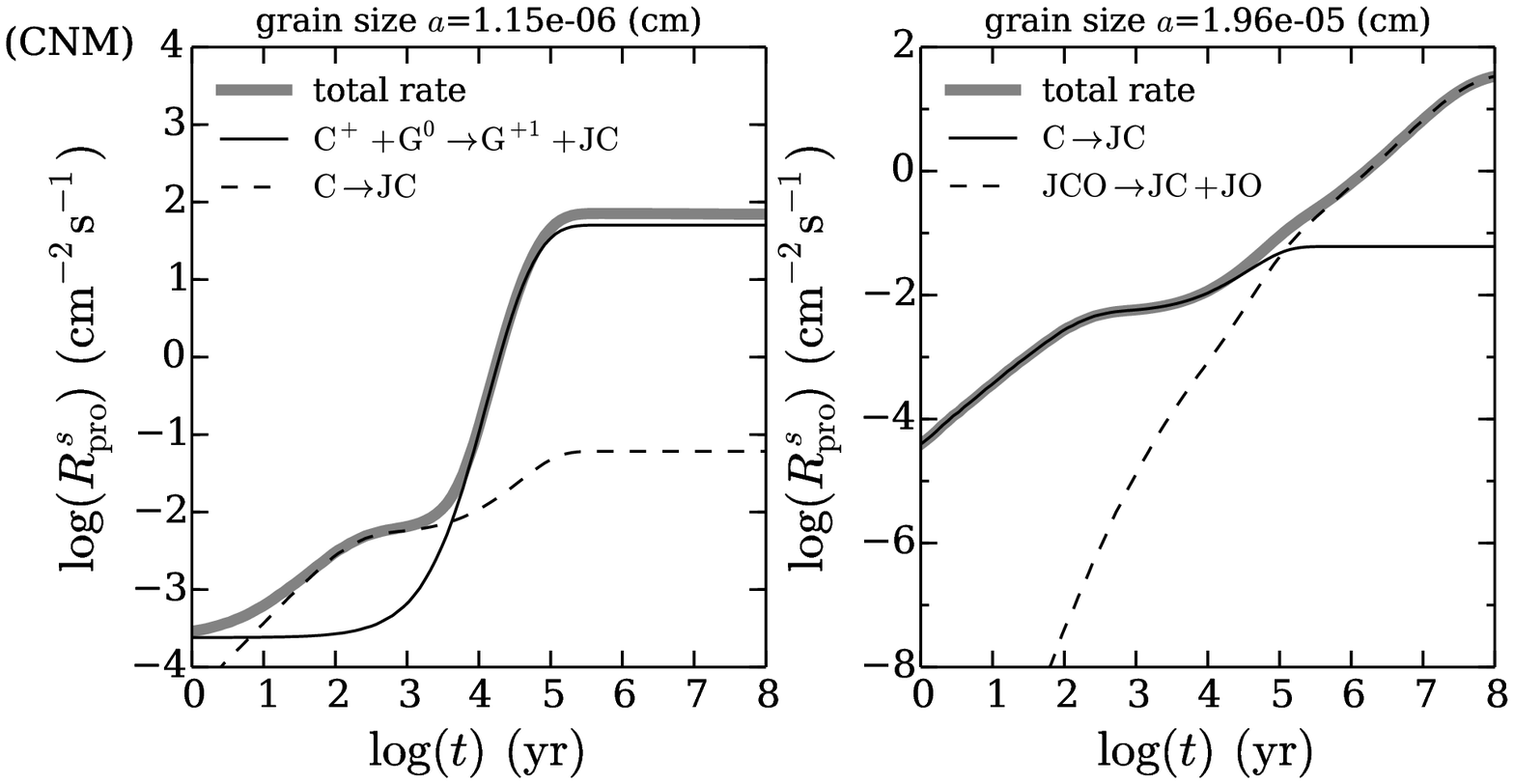}
\caption{JC reaction rate tracing (RRT) diagram of the production rate $R^s_{\rm pro}(a)$ for the surface densities on the smallest (left panels) and biggest (right panels) grains. The upper and lower rows are for the DC and CNM models, respectively. Only the most important contributing reactions to the total production rates are shown.}
\label{DC_CNM_RRT_JC}
\end{figure}
\begin{figure}
\centering
\includegraphics[scale=0.41]{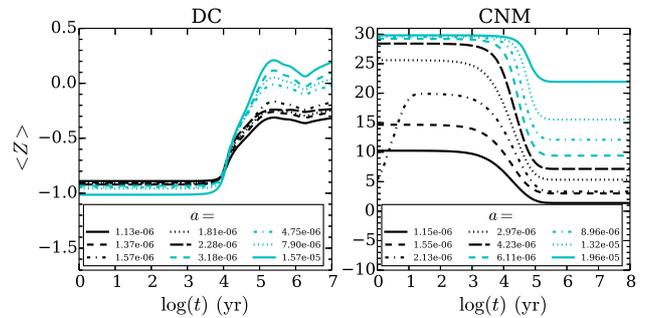}
\caption{The averaged charge evolution tracks for grains in the DC (left panel) and CNM (right panel) models.}
\label{charge_dc_cnm}
\end{figure}
To explain the chemical trends in our models, we apply the Reaction Rate Tracing (RRT) method defined by GHY to some selected surface species. Instead of using the reaction rate $R(a)$ for volume densities, we use the reaction rate $R^s(a)=R(a)/\left[ 4\pi a^2 n_{\rm d}(a)\right]$ for surface densities. We only show the most important chemical processes contributing to the selected species. Here, we also take JC as an example to present the RRT diagram of production rate in Fig.~\ref{DC_CNM_RRT_JC}. 

The RRT diagrams of JC in the DC model (upper panels in Fig.~\ref{DC_CNM_RRT_JC}) demonstrate that the accretions of C$^+$ onto grain G$^{-1}$ (G$^Z$ denotes grains with charge number $Z$) are the leading processes on both the smallest (left panel) and biggest (right panel) grains at $t < 10^4\yr$. We note that the surface density production rate on the smallest grains are larger than that on the biggest grains due to the bigger Coulomb factor for smaller grains, which results in the enhancement of areal density on smaller grains by factors up to $\simali$1 orders of magnitude (see in Fig.~\ref{dc_cnm_i_density}). At $t>10^4\yr$, the C$^+$ abundance drops and the C abundance becomes very high due to the accretion and recombination with electron, and thus the C atom accretion becomes the leading processes on all grains. Because the neutral accretion rate coefficients are not dependent on grain sizes, JC shows the same area densities on all grains in this period in the upper left panel of Fig.~\ref{dc_cnm_i_density}. This is also evidenced by the grain charge evolution, see the left panel of Fig.~\ref{charge_dc_cnm}, which shows that the grains are negatively charged at $t<10^5\yr$, after that they become nearly neutral and this weakens the coulomb attraction for C$^+$.

For JC in the CNM model, the RRT diagrams (lower panels in Fig.~\ref{DC_CNM_RRT_JC}) show that the leading formation process on the smallest grains is the accretion of C$^+$ onto neutral grains (G$^{0}$), except at early times $t<5000\yr$ when it is dominated by accretion of neutral C. For the largest grains, the neutral processes (i.e., the C atom accretion at $t<10^5\yr$ and the surface reaction between JC and JO after that) play the most important role in making JC. This is due to that the smallest and biggest grains possess different average charges of $\sim +1{\rm e}$ and $\sim +22{\rm e}$ respectively (see the right panel of Fig.~\ref{charge_dc_cnm}), which results in efficient C$^+$ accretion on the smallest grains to recover its leading role at $t>5\times10^3\yr$, while the accretion of C$^+$ is always unimportant on the biggest grains due to the high Coulomb repulsive force. 

We note that in the simulations reported above the thermal balance temperatures are adopted for all grains. However, tiny grains smaller than about $0.01\micron$ may experience considerable temperature fluctuations upon single photon heating events \citep{Draine2001I}, which may introduce additional chemical differentiation over grain sizes. This is particularly true in our CNM model in which the radiation field is strong. A comparison of the grain temperature fluctuation time scales with the time scales of gas accretion onto grains, ice thermal desorption and surface chemical reactions in our CNM model shows that (1) the grain temperature fluctuation time scales are always much shorter than the gas accretion time scales; (2) a sudden rise of the grain temperature greatly speeds up thermal desorption \citep[see also][]{Aann1979} and surface reactions of volatile surface species. Thus, we speculate that the grain temperature fluctuation tends to clear out volatile species from the  surface of smaller grain through thermal desorption or converting them into more inert heavy species by thermally enhanced surface reactions. This will be investigated in a subsequent paper.

\section{Conclusions}
\label{conclusions}
We have demonstrated that, when grain charging is handled properly, the inclusion of ion accretion onto grains and the size-dependency of grain temperature will result in a significant differentiation of areal densities of surface species over grain sizes. This renders the assumption of a constant surface density over grain sizes in gas-grain chemical modeling no longer valid. Our gas-grain chemical simulations of the DC and CNM models with a simplified chemical network have not only confirmed the differentiation effects, but also showed that shifting from a single grain size to a more realistic size distribution can incur a large variation of surface abundances by up to $\simali$2--4 orders of magnitude, even though the total grain surface area is kept unchanged. We thus conclude that the chemical differentiation across grain sizes results in measurable changes of the abundances of icy species with respect to that of the single grain-size model.

Our results have built up a link between ice formation and grain size distribution and grain charging processes. Together with the chemical effects of the variation of the total grain surface area in literature, the dust grain size distribution should have significant consequences upon hot core chemistry in which the ices will be differentially sublimated from grains of different sizes into the gas phase. The chemical effects may also open a new possiblity for investigating the dust grain size distributions through observations of ice absorption bands in the infrared.

\section*{Acknowledgements}
We thank the anonymous referee for his/her very helpful comments. JH thanks the support of the NSFC grant No. 11173056 and the academic leader talent plan of Guan-du district of Kunming. AL is supported by NSF AST-1311804 and NNX13AE63G. We also thank Dr. Hiroyuki Hirashita for generously offering his grain size distribution data.



\bibliographystyle{mnras}
\bibliography{letter} 








\bsp	
\label{lastpage}
\end{document}